\documentclass[12pt, draftclsnofoot, onecolumn]{IEEEtran}
\usepackage{amsmath,amssymb}
\usepackage{amscd,amsfonts}
\usepackage{color,graphicx}
\usepackage{epstopdf}
\usepackage{subfigure}
\usepackage{algorithm, algorithmic}
\usepackage{mathtools}
\allowdisplaybreaks
\newtheorem{theorem}{Theorem}
\begin{document}

%=========================================================================%
\title{Beamforming for Full-Duplex Multiuser MIMO Systems}
\author{\IEEEauthorblockN{Jinwoo Kim, \textit{Student Member, IEEE}, Wan Choi,
\textit{Senior Member, IEEE}, and Hyuncheol Park, \textit{Senior Member, IEEE},}
\thanks{J. ~Kim, W.~Choi, and H.~Park are with School of Electrical
    Engineering, Korea Advanced Institute of Science and Technology
    (KAIST), Daejeon 305-701, Korea (e-mail: jwkim901@kaist.ac.kr, wchoi@kaist.edu,
    hcpark@kaist.ac.kr).}
    }
%=========================================================================%
\maketitle
\vspace{-0.8in}

%=========================================================================%
\begin{abstract}
We solve a sum rate maximization problem of full-duplex (FD) multiuser multiple-input multiple-output (MU-MIMO) systems. Since additional self-interference (SI) in the uplink channel and co-channel interference (CCI) in the downlink channel are coupled in FD communication, the downlink and uplink multiuser beamforming vectors are required to be jointly designed. However, the joint optimization problem is non-convex and hard to solve due to the coupled effect. To properly address the coupled design issue, we reformulate the problem into an equivalent uplink channel problem, using the uplink and downlink channel duality known as MAC-BC duality. Then, using minorization maximization (MM) algorithm based on an affine approximation, we obtain a solution for the reformulated problem. In addition, without any approximation and thus performance degradation, we develop an alternating algorithm based on iterative water-filling (IWF) to solve the non-convex problem. The proposed algorithms warrant fast convergence and low computational complexity.

\end{abstract}
%=========================================================================%

%=========================================================================%
%=========================================================================%
\newpage
%=========================================================================%
%                                                                     %
%=========================================================================%
\section{Introduction}
%=========================================================================%

% Necessity of full-duplex, definition and operation challenges
The rapid proliferation of wireless devices and related service results in the recent exploding demand on extra frequency bands. Because  available frequency bands are limited,
spectral efficiency becomes a key design criterion of wireless communication systems. Full-duplex (FD) operation has received a great attention since theoretically it is able to double spectral efficiency compared to half-duplex (HD) operation. In FD operation, presuming the self-interference (SI) from the transmitted signal is properly suppressed,
simultaneous transmission and reception are allowed in the same frequency band. To make FD communication viable,  SI cancellation techniques are essential, and fortunately
the recent advancement of SI cancellation techniques sheds light on practical feasibility of FD communication \cite{Choi10Achieving}--\cite{Duarte12Experiment}. Some experimental results based on the advanced SI suppression techniques demonstrated a possibility of FD communication in real environments \cite{Choi10Achieving}\cite{Duarte10Full}. In \cite{Jain11Practical}, an adaptive cancellation scheme was proposed to overcome some practical limitations of the previous SI cancellation schemes. A combination of passive SI suppression and active SI cancellation was shown to achieve 74 dB suppression of SI  on average \cite{Duarte12Experiment}.

% Beamforming (SU-MIMO, MU-MIMO), previous research
FD communication can be leveraged by beamforming with multiple antennas  \cite{Sabharwal14Inband}. There exists residual SI due to imperfect SI suppression, and beamforming can be exploited to address the residual SI. In particular, to maximize sum rate of a FD system, beamforming has to balance between residual SI suppression and information transfer. In bi-directional communications, an iterative precoding technique based on sequential convex programming (SCP) was developed to balance sum rate maximization and SI suppression, using appropriate weighting factors \cite{Huberman13Self, Huberman14Sequential}. In FD multiuser network where a FD base station (BS) concurrently serves uplink HD users and downlink HD users in the same frequency band, the base station suffers from SI and the downlink users are interfered by the transmitted signals of the uplink users, i.e., co-channel interference (CCI). If the downlink transmit power increases to combat CCI perceived at the downlink users, SI also increases at the base station and thus the uplink sum rate decreases. On the other hand, if the uplink users increase transmit power against SI at the base station, CCI increases at the downlink users. Thus, in the FD multiuser network, transmission strategies at the base station and the uplink users are coupled and have to be designed to address both CCI and SI simultaneously, which poses a jointly coupled optimization problem. In  \cite{Nguyen14On}, as a simplified problem for the FD multiuser system, single antenna users were considered when the base station performs linear beamforming for the downlink users and non-linear multiuser detection, i.e., minimum mean-square-error successive interference cancellation (MMSE-SIC),  for the uplink users. Then, the downlink beamformer design problem was formulated as a rank-1 constrained optimization problem and suboptimal solutions were presented based on rank relaxation and approximations.
Multiple antenna in full duplex multiuser systems (FD MU-MIMO)  were studied in \cite{Nguyen12Transmission}--\cite{Huberman15MIMO}.  Uplink beamformer design and downlink power allocation addressing SI at the base station (BS) was studied in \cite{Nguyen12Transmission, Nguyen13Precoding}. However, CCI was discarded and  zero-forcing (ZF) downlink beamforming at BS was assumed for simplicity, albeit its suboptimality.  In \cite{Yin13Full}, both CCI and SI were considered but with the assumption of large scale MIMO, ZF downlink beamforming was simply used for SI suppression while treating CCI as a background noise. The authors of \cite{Huberman14Full, Huberman15MIMO} addressed SI and CCI simultaneously based on SCP algorithms in linear beamformer design.

% Objective of this paper
In this paper, when the uplink users and the downlink users have multiple antennas in the FD multiuser system, we explore novel transmission strategies at the base station and the uplink users in terms of maximizing sum rate of the uplink and downlink users. To this end, we formulate a joint beamformer design problem to maximize sum rate, i.e., a  joint transmit covariance matrix design problem, modeling the coupled effects between SI and CCI. However, the optimization problem is non-convex and not easy to find the optimal transmit covariance matrices due to the coupled SI and CCI. To circumvent this difficulty, we exploit the duality between broadcast channel (BC) and multiple access channel (MAC) \cite{Vishwanath03Duality} and reformulate the sum rate maximization problem as an equivalent optimization problem for MAC. Although the reformulated problem is still non-convex, the objective is represented as a difference of two concave functions and then, using the minorization maximization (MM) algorithm based an affine approximation, the objective can be approximated as a concave function. Accordingly, we solve the problem with disciplined convex programming (DCP) using cvx program \cite{CVX13}. In addition, without any approximation of the objective function and thus performance degradation, we develop an alternating iterative water-filling (IWF) algorithm to solve the non-convex problem. The proposed algorithm is based on the iterative water-filling algorithm \cite{Yu04Iterative, Jindal05Sum} which is known to provide the optimal transmit covariance matrices for MAC. The proposed algorithms ensure fast convergence and low computational complexity.
Compared to \cite{Huberman14Full} and \cite{Huberman15MIMO} which address SI and CCI simultaneously as in our paper, the design approach differs; in \cite{Huberman14Full} and \cite{Huberman15MIMO}, uplink and downlink  linear beamformers were developed with the SCAMP algorithm and the cvx solver, both of which are based on the SCP approach. On the contrary,  our non-linear beamformer design is based on dirty paper coding in downlink and MMSE-SIC in uplink, which are known as capacity achieving schemes  in downlink MU-MIMO and uplink MU-MIMO, respectively, and the transmit covariance matrices are found with the proposed algorithms based on the MAC-BC duality. The proposed MM algorithm differs from the algorithms in \cite{Huberman14Full} and \cite{Huberman15MIMO}  in the respect that it is used for non-linear beamforming, although it is also based on the SCP approach. Moreover, since the DC-based algorithms using affine approximations can suffer from information loss, we proposed the alternating iterative water-filling algorithm  enabled by the MAC-BC duality, which does not rely on  cvx solver requiring  long computational time.

The remainder of this paper is organized as follows.
In Section \ref{Sec:Sys_model}, we describe the system and channel model and then formulate the design problem.
The proposed iterative beamforming algorithms are proposed  in  Section \ref{Sec:BF_design}. Numerical results are presented in
Section \ref{Sec:Results}. Finally, conclusions are drawn in Section \ref{Sec:Conclusion}.

%=========================================================================%
%                                                                     %
%=========================================================================%
\section{System Model and Problem Formulation}\label{Sec:Sys_model}
%=========================================================================%

%=========================================================================%
\begin{figure}[!t]
    \begin{center}
\centering
\includegraphics[width=9cm]{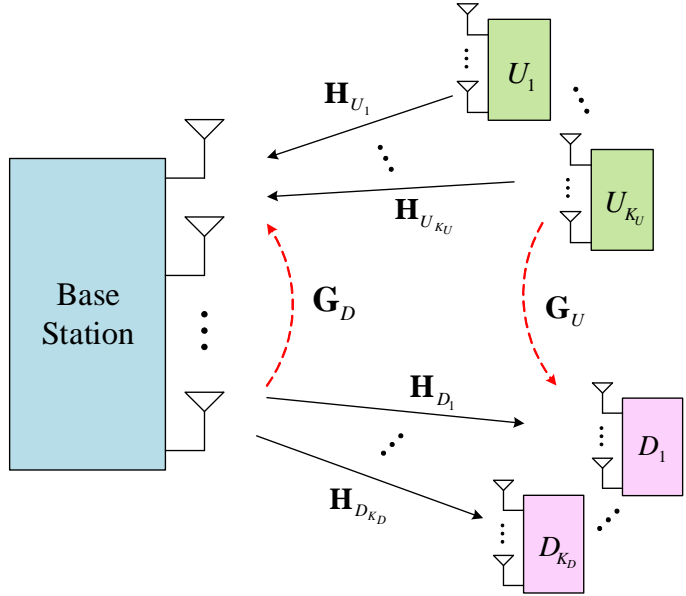}
    \caption{A full-duplex MU-MIMO system model}
    \label{FD_model}
    \end{center}
\end{figure}
%=========================================================================%

\subsection{System and Channel Model}
We consider a single cell FD MU-MIMO system as shown in Fig. \ref{FD_model}, where a FD base station (BS) with $M$ antennas concurrently serves $K_{U}$ uplink users and $K_{D}$ downlink users with $N$ antennas each.
Since the BS transmits and receives simultaneously in the same frequency band, the transmitted signal unavoidably interferes with the received signals from the uplink users, which is called self-interference (SI). Even with recent advanced SI cancellation techniques, there exists residual SI\footnote{Since residual SI results from limited cancellation capability due to channel estimation error, analog-digital converter (ADC) resolution, and so on, we do not consider perfect CSI of the SI channel, while perfect CSI of user channels is assumed to focus on the effect of residual SI on beamforming in FD MU-MIMO systems.} due to imperfect SI cancellation. Moreover, the downlink users suffer from co-channel interference (CCI) caused by the uplink users.

Let $\mathbf{H}_{U_{i}} \in \mathbb{C}^{M \times N}$ and $\mathbf{H}_{D_{j}} \in \mathbb{C}^{N \times M}$  be the channel matrices from uplink user $U_{i}$ to the BS  and from the BS to downlink user $D_{j}$, respectively, each element of which is an independent and identically distributed (i.i.d.) complex Gaussian random variable with zero mean and unit variance.
$\mathbf{G}_{D} \in \mathbb{C}^{M \times M}$ represents the SI channel matrix which typically models the residual SI after SI cancellation. The channel matrix of CCI is given by
$\mathbf{G}_{U}=[(\mathbf{G}_{U}^{D_{1}})^{H}~\cdots~(\mathbf{G}_{U}^{D_{K_{D}}})^{H}]^{H}$, where $\mathbf{G}_{U}^{D_{j}}=[\mathbf{G}_{U_{1}}^{D_{j}}~\cdots~\mathbf{G}_{U_{K_{U}}}^{D_{j}}]$ and $\mathbf{G}_{U_{i}}^{D_{j}} \in \mathbb{C}^{N \times N}$ represents the CCI from $U_{i}$ to $D_{j}$. The entries of SI and CCI channel matrices are assumed to be i.i.d. complex Gaussian random variables with zero mean and variance $\sigma_{\mathrm{SI}}^{2}$ and $\sigma_{\mathrm{CCI}}^{2}$, respectively.

%=========================================================================%
\begin{figure*}[!t]
\centering
    \subfigure[]{
        \includegraphics[width=9cm]{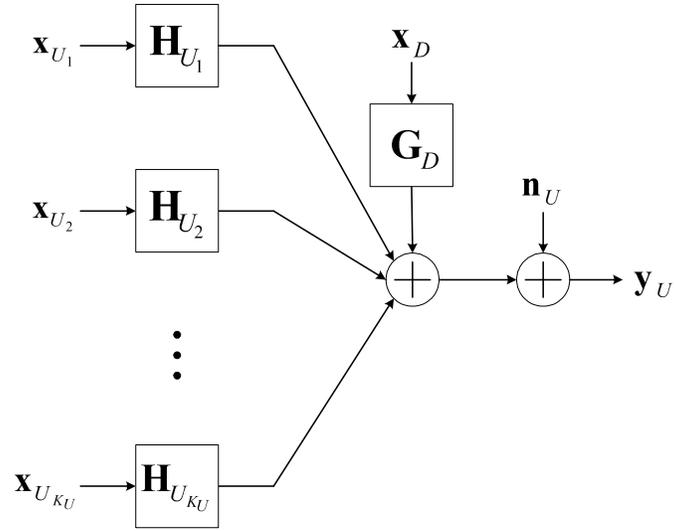}
        \label{UL_model}
    }
    \subfigure[]{
        \includegraphics[width=9cm]{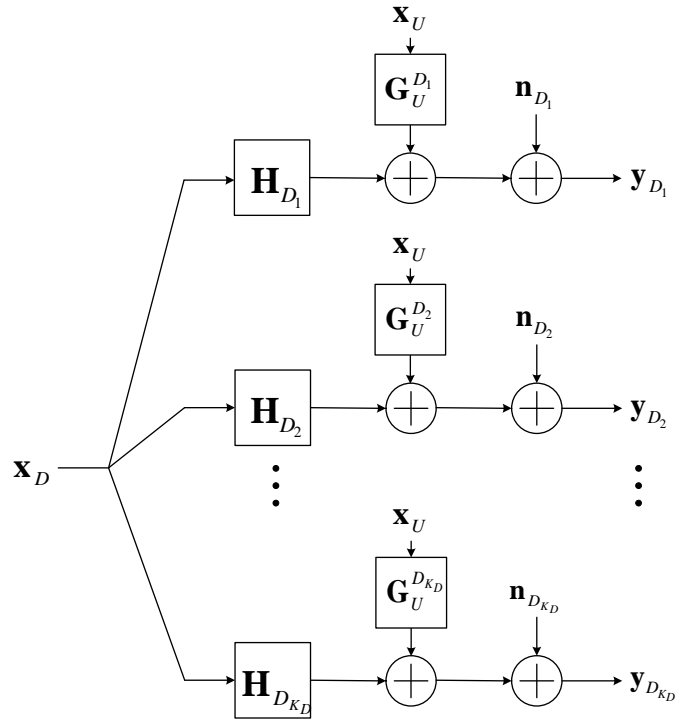}
        \label{DL_model}
    }
\caption{(a) Uplink channel with self-interference
(b) Downlink channel with co-channel interference}
\label{UL_DL_model}
\end{figure*}
%=========================================================================%
The received signal at the BS is represented as
\begin{equation}\label{UL_RX}
    \mathbf{y}_{U} = \sum_{i=1}^{K_{U}}\mathbf{H}_{U_{i}}\mathbf{x}_{U_{i}} + \sum_{j=1}^{K_{D}}\mathbf{G}_{D}\mathbf{x}_{D_{j}} + \mathbf{n}_{U}
\end{equation}
where $U_{i}$ and $D_{j}$ represent the index of the $i$-th user in the uplink channel and the index of the $j$-th user in the downlink channel, respectively, $\mathbf{x}_{U_{i}} \in \mathbb{C}^{N \times 1}$ is the transmitted signal vector of uplink user $U_{i}$, $\mathbf{x}_{D_{j}} \in \mathbb{C}^{M \times 1}$ is the transmitted signal vector from the BS to the $j$-th downlink user, and $\mathbf{n}_{U} \in \mathbb{C}^{M \times 1}$ is the additive white Gaussian noise (AWGN) vector with zero mean and covariance $\mathbb{E}(\mathbf{n}_{U}\mathbf{n}_{U}^{H})=\mathbf{I}_{M}$.

On the other hand, the received signal at $D_{j}$ is represented as
\begin{equation}\label{DL_RX}
    \mathbf{y}_{D_{j}} = \mathbf{H}_{D_{j}}\mathbf{x}_{D}
    + \sum_{i=1}^{K_{U}}\mathbf{G}_{U_{i}}^{D_{j}}\mathbf{x}_{U_{i}} + \mathbf{n}_{D_{j}}, \quad j=1,...,K_{D}
\end{equation}
where $\mathbf{n}_{D_{j}} \in \mathbb{C}^{N \times 1}$ is the AWGN vector with zero mean and covariance $\mathbb{E}(\mathbf{n}_{D_{j}}\mathbf{n}_{D_{j}}^{H})=\mathbf{I}_{N}$.
Defining $\mathbf{x}_{U}=[\mathbf{x}_{U_{1}}^{T}~\cdots~\mathbf{x}_{U_{K_{U}}}^{T}]^{T}$, the FD MU-MIMO system can be decomposed into an uplink channel with SI and a downlink channel with CCI, as shown in Fig.~\ref{UL_model} and Fig.~\ref{DL_model}, respectively, although they are still coupled each other.

%=========================================================================%
\subsection{Problem Formulation}
%=========================================================================%
Without loss of generality, we can assume that the uplink users are decoded in order, from $U_{1}$ to $U_{K_{U}}$. Using MMSE-SIC, the achievable rate of uplink user $U_{i}$ is obtained as
\begin{equation}\label{R_UL}
    R_{U_{i}} = \log{\frac{\left| \mathbf{I}+\sum_{j=1}^{K_{D}}\mathbf{G}_{D}\mathbf{Q}_{D_{j}}\mathbf{G}_{D}^{H}
    +\sum_{k=i}^{K_{U}}\mathbf{H}_{U_{k}}\mathbf{Q}_{U_{k}}\mathbf{H}_{U_{k}}^{H}\right|}
    {\left| \mathbf{I}+\sum_{j=1}^{K_{D}}\mathbf{G}_{D}\mathbf{Q}_{D_{j}}\mathbf{G}_{D}^{H}
    +\sum_{k=i+1}^{K_{U}}\mathbf{H}_{U_{k}}\mathbf{Q}_{U_{k}}\mathbf{H}_{U_{k}}^{H}\right|}}, \quad i=1,\ldots,K_{U},
\end{equation}
where $\mathbf{Q}_{U_{i}}$ and $\mathbf{Q}_{D_{j}}$ denote the uplink transmit covariance matrix
of $U_{i}$ and the downlink transmit covariance matrix of $D_{j}$, respectively, and $|\cdot|$ is the determinant operator.

In the downlink channel, the achievable rate of user $\pi_{D}(j)$ is given by
\begin{equation}\label{R_DL}
    R_{\pi_{D}(j)} = \log{\frac{\left| \mathbf{I}+\sum_{i=1}^{K_{U}}\mathbf{G}_{U_{i}}^{\pi_{D}(j)}\mathbf{Q}_{U_{i}}(\mathbf{G}_{U_{i}}^{\pi_{D}(j)})^{H}
    +\mathbf{H}_{\pi_{D}(j)}(\sum_{k=j}^{K_{D}}\mathbf{Q}_{\pi_{D}(k)})\mathbf{H}_{\pi_{D}(j)}^{H}\right|}
    {\left| \mathbf{I}+\sum_{i=1}^{K_{U}}\mathbf{G}_{U_{i}}^{\pi_{D}(j)}\mathbf{Q}_{U_{i}}(\mathbf{G}_{U_{i}}^{\pi_{D}(j)})^{H}
    +\mathbf{H}_{\pi_{D}(j)}(\sum_{k=j+1}^{K_{D}}\mathbf{Q}_{\pi_{D}(k)})\mathbf{H}_{\pi_{D}(j)}^{H}\right|}}, \quad j=1,\ldots,K_{D},
\end{equation}
where $\pi_{D}(j)$ denotes the index of the $j$-th encoded user based on dirty paper coding (DPC) \cite{Costa83Writing}.
We assume that downlink users are encoded in order, from $D_{K_{D}}$ to $D_{1}$, $\{ \pi_{D}(j) \}_{j=1}^{K_{D}}=\{D_{K_{D}},...,D_{1}\}$. Then, the sum rate
maximization problem for the system is formulated as
\begin{align}\label{Orig_Sumrate}
    \underset{ \{\mathbf{Q}_{U_{i}}\}, \{\mathbf{Q}_{D_{j}}\} }{\mathrm{max}} \quad & \sum_{i=1}^{K_{U}}R_{U_{i}} + \sum_{j=1}^{K_{D}}R_{D_{j}} \nonumber \\
    \mathrm{subject~to} \quad & \mathrm{Tr} \left(\mathbf{Q}_{U_{i}}\right) \leq P_{U_{i}}, \quad i=1,\ldots,K_{U} \nonumber \\
                                & \sum_{j=1}^{K_{D}}\mathrm{Tr} \left( \mathbf{Q}_{D_{j}} \right) \leq P_{D}  \nonumber  \\
                                & \mathbf{Q}_{U_{i}} \succeq 0,~\mathbf{Q}_{D_{j}} \succeq 0, \quad i=1,\ldots,K_{U},~j=1,\ldots,K_{D},
\end{align}
where $\succeq$ denotes that the matrix on the left side of which is a positive semi-definite matrix. $P_{U_{i}}$ is the maximum power for $U_{i}$, and $P_{D}$ is the maximum power of the BS. The objective function is non-convex in nature, and the SI term in the uplink sum rate and the CCI term in the downlink sum rate are coupled, which makes the problem further difficult to solve.

%=========================================================================%
%                                                                     %
%=========================================================================%
\section{Beamformer Design}\label{Sec:BF_design}
%=========================================================================
%
%=========================================================================%
\subsection{MAC-BC Duality}
%=========================================================================%
In the downlink channel, the interference from uplink users (CCI) plus additive white  Gaussian noise at downlink user $D_{j}$ is represented as
\begin{equation}
 \mathbf{\tilde{n}}_{D_{j}} = \sum_{i=1}^{K_{U}}\mathbf{G}_{U_{i}}^{D_{j}}\mathbf{x}_{U_{i}} + \mathbf{n}_{D_{j}}
\end{equation} which, assuming Gaussian signaling, becomes a colored Gaussian noise with zero mean and the covariance matrix given by
\begin{align}
\mathbb{E}\left[ \mathbf{\tilde{n}}_{D_{j}}\mathbf{\tilde{n}}_{D_{j}}^{H} \right]&=\mathbf{W}_{D_{j}} \nonumber \\
&=\mathbf{I}+\sum_{i=1}^{K_{U}}\mathbf{G}_{U_{i}}^{D_{j}}\mathbf{Q}_{U_{i}}\left(\mathbf{G}_{U_{i}}^{D_{j}}\right)^{H}.
\end{align}

Applying a whitening filter at receiver $D_{j}$, we obtain the received signal as
\begin{align}
\mathbf{\bar{y}}_{D_{j}} &= \mathbf{W}_{D_{j}}^{-1/2}\mathbf{y}_{D_{j}} \nonumber \\
&= \mathbf{W}_{D_{j}}^{-1/2}\mathbf{H}_{D_{j}}\mathbf{x}_{D_{j}} +\mathbf{W}_{D_{j}}^{-1/2}\mathbf{\tilde{n}}_{D_{j}} \nonumber \\
&= \mathbf{\bar{H}}_{D_{j}}\mathbf{x}_{D_{j}} + \mathbf{\bar{n}}_{D_{j}},
\end{align}
where $\mathbf{\bar{H}}_{D_{j}}$ is the effective channel for downlink user $D_{j}$. Now, to circumvent the difficulty involved in the non-convex optimization problem, we use the duality between MAC and BC for the effective downlink channels $\{\mathbf{\bar{H}}_{D_{j}}\}$ after whitening.
% Duality
%=========================================================================%
\begin{figure}[t]
    \begin{center}
\centering
\includegraphics[width=9cm]{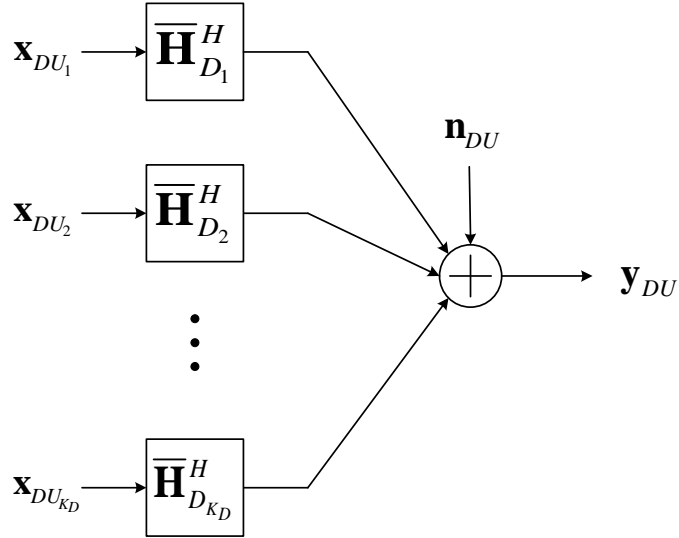}
    \caption{Dual uplink channel with co-channel interference}
    \label{Dual_UL_model}
    \end{center}
\end{figure}
%=========================================================================%
Based on the MAC-BC duality, the received signal in the dual uplink channel can be written as
\begin{equation}\label{DUL_RX}
    \mathbf{y}_{DU} = \sum_{j=1}^{K_{D}}\mathbf{\bar{H}}_{D_{j}}^{H}\mathbf{x}_{DU_{j}} + \mathbf{n}_{DU}
\end{equation}
where $DU_{j}$ represents the index of the $j$-th user in the dual uplink channel, $\mathbf{x}_{DU_{j}} \in \mathbb{C}^{N \times 1}$ is the transmit signal in the dual uplink channel,
and $\mathbf{n}_{DU} \in \mathbb{C}^{M \times 1}$ is the additive white Gaussian noise vector, as in Fig.~\ref{Dual_UL_model}.
With the dual uplink channel, the sum rates of the uplink and the downlink channels are represented, respectively, as
\begin{equation}\label{R_UL}
    R_{U} = \log{\frac{\left| \mathbf{I}+\sum_{j=1}^{K_{D}}\mathbf{G}_{D}\mathbf{Q}_{D_{j}}\mathbf{G}_{D}^{H} + \sum_{i=1}^{K_{U}}\mathbf{H}_{U_{i}}\mathbf{Q}_{U_{i}}\mathbf{H}_{U_{i}}^{H}\right|}
    {\left| \mathbf{I}+\sum_{j=1}^{K_{D}}\mathbf{G}_{D}\mathbf{Q}_{D_{j}}\mathbf{G}_{D}^{H}\right|}}
\end{equation}
and
\begin{equation}\label{R_DUL}
    R_{DU} = \log\left| \mathbf{I} + \sum_{j=1}^{K_{D}}\mathbf{\bar{H}}_{D_{j}}^{H}\mathbf{Q}_{DU_{j}}\mathbf{\bar{H}}_{D_{j}}\right|,
\end{equation}
where $\mathbf{Q}_{DU_{j}}$ is the transmit covariance matrix of dual uplink user $DU_{j}$. From the MAC-BC duality \cite{Vishwanath03Duality}, $\mathbf{Q}_{DU_{j}}$ and $\mathbf{Q}_{D_{j}}$ have the following relationship:
\begin{align}\label{MACtoBC}
\mathbf{Q}_{D_{j}} &= \mathbf{B}_{D_{j}}^{-1/2}\overline{\mathbf{A}_{D_{j}}^{1/2}\mathbf{Q}_{DU_{j}}\mathbf{A}_{D_{j}}^{1/2}}\mathbf{B}_{D_{j}}^{-1/2} \nonumber \\
		&= \mathbf{B}_{D_{j}}^{-1/2}\mathbf{F}_{D_{j}}\bar{\mathbf{F}}_{D_{j}}^{H}\mathbf{A}_{D_{j}}^{1/2}\mathbf{Q}_{DU_{j}}\mathbf{A}_{D_{j}}^{1/2}\bar{\mathbf{F}}_{D_{j}}\mathbf{F}_{D_{j}}^{H}\mathbf{B}_{D_{j}}^{-1/2}
\end{align}
where $ \mathbf{A}_{D_{j}} = \mathbf{I} + \mathbf{\bar{H}}_{D_{j}}\left(\sum_{k=1}^{j-1}\mathbf{Q}_{D_{k}} \right)\mathbf{\bar{H}}_{D_{j}}^{H}$, $
\mathbf{B}_{D_{j}} = \mathbf{I} + \sum_{k=j+1}^{K_{D}}\mathbf{\bar{H}}_{D_{k}}^{H}\mathbf{Q}_{DU_{k}}\mathbf{\bar{H}}_{D_{k}}$, and
$\mathbf{F}_{D_{j}}\mathbf{\Lambda}_{D_{j}}\bar{\mathbf{F}}_{D_{j}}^{H}$ is the singular value decomposition of the effective channel such that
\begin{align}
\mathbf{B}_{D_{j}}^{-1/2}\mathbf{\bar{H}}_{D_{j}}^{H}\mathbf{A}_{D_{j}}^{-1/2} = \mathbf{F}_{D_{j}}\mathbf{\Lambda}_{D_{j}}\bar{\mathbf{F}}_{D_{j}}^{H}.
\end{align}

 Using (\ref{R_UL}) and (\ref{R_DUL}), we can reformulate the sum rate maximization problem as
% Optimization problem using duality
\begin{align}\label{New_Sumrate}
     \underset{ \{\mathbf{Q}_{U_{i}}\}, \{\mathbf{Q}_{DU_{j}}\} }{\mathrm{max}} \quad & R_{U} + R_{DU}
    % =  \log{\frac{\left| \mathbf{I}+\sum_{j=1}^{K_{D}}\mathbf{\bar{G}}_{D_{j}}\mathbf{Q}_{DU_{j}}\mathbf{\bar{G}}_{D_{j}}^{H}
    %+\sum_{i=1}^{K_{U}}\mathbf{H}_{U_{i}}\mathbf{Q}_{U_{i}}\mathbf{H}_{U_{i}}^{H}\right|}
    %{\left| \mathbf{I}+\sum_{j=1}^{K_{D}}\mathbf{\bar{G}}_{D_{j}}\mathbf{Q}_{DU_{j}}\mathbf{\bar{G}}_{D_{j}}^{H}\right|}} \nonumber \\
    %& \quad \quad \quad \quad ~~~~ +\log{\frac{\left| \mathbf{I}+\sum_{i=1}^{K_{U}}\mathbf{\bar{G}}_{U_{i}}\mathbf{Q}_{U_{i}}\mathbf{\bar{G}}_{U_{i}}^{H}
    %+\sum_{j=1}^{K_{D}}\mathbf{H}_{D_{j}}^{H}\mathbf{Q}_{DU_{j}}\mathbf{H}_{D_{j}}\right|}
    %{\left| \mathbf{I}+\sum_{i=1}^{K_{U}}\mathbf{\bar{G}}_{U_{i}}\mathbf{Q}_{U_{i}}\mathbf{\bar{G}}_{U_{i}}^{H}\right|}}
    \nonumber \\
    \mathrm{subject~to} \quad & \mathrm{Tr} \left(\mathbf{Q}_{U_{i}}\right) \leq P_{U_{i}}, \quad \mathbf{Q}_{U_{i}} \succeq 0, ~\quad i=1,\ldots,K_{U} \nonumber \\
                                & \sum_{j=1}^{K_{D}}\mathrm{Tr} \left( \mathbf{Q}_{DU_{j}} \right) \leq P_{D},  \quad \mathbf{Q}_{DU_{j}} \succeq 0, \quad j=1,\ldots,K_{D}.
\end{align}

%=========================================================================%
\subsection{Minorization Maximization (MM) algorithm}
%=========================================================================%
From the reformulated problem in (\ref{New_Sumrate}), which is still non-convex, the objective can be represented as a difference of concave functions (DC) as
\begin{align}\label{Diff_Con}
    f\left(\mathbf{Q}_{U},\mathbf{Q}_{DU}\right) &= R_{U} + R_{DU} \nonumber \\
    &=  g\left(\mathbf{Q}_{U},\mathbf{Q}_{DU}\right) - h\left(\mathbf{Q}_{DU}\right)
\end{align}
where
$$g\left(\mathbf{Q}_{U},\mathbf{Q}_{DU}\right) = \log \left|\mathbf{I} + \sum_{j=1}^{K_{D}}\mathbf{\bar{G}}_{D_{j}}\mathbf{Q}_{DU_{j}}\mathbf{\bar{G}}_{D_{j}}^{H}
+ \sum_{i=1}^{K_{U}}\mathbf{H}_{U_{i}}\mathbf{Q}_{U_{i}}\mathbf{H}_{U_{i}}^{H}\right| + \log \left|\mathbf{I} + \sum_{j=1}^{K_{D}}\mathbf{\bar{H}}_{D_{j}}^{H}\mathbf{Q}_{DU_{j}}\mathbf{\bar{H}}_{D_{j}}\right|$$
and $$h\left(\mathbf{Q}_{DU}\right)= \log \left|\mathbf{I} + \sum_{j=1}^{K_{D}}\mathbf{\bar{G}}_{D_{j}}\mathbf{Q}_{DU_{j}}\mathbf{\bar{G}}_{D_{j}}^{H}\right|.$$
From the MAC-BC transformation, the SI channel is rewritten as $\mathbf{\bar{G}}_{D_{j}}=\mathbf{G}_{D}\mathbf{B}_{D_{j}}^{-1/2}\mathbf{F}_{D_{j}}\mathbf{\bar{F}}_{D_{j}}^{H}\mathbf{A}_{D_{j}}^{1/2}$ where $
    \mathbf{A}_{D_{j}} = \mathbf{I} + \mathbf{\bar{H}}_{D_{j}}\left(\sum_{k=1}^{j-1}\mathbf{Q}_{D_{k}} \right)\mathbf{\bar{H}}_{D_{j}}^{H}$, $
\mathbf{B}_{D_{j}} = \mathbf{I} + \sum_{k=j+1}^{K_{D}}\mathbf{\bar{H}}_{D_{k}}^{H}\mathbf{Q}_{DU_{k}}\mathbf{\bar{H}}_{D_{k}}$, and $\mathbf{B}_{D_{j}}^{-1/2}\mathbf{\bar{H}}_{D_{j}}^{H}\mathbf{A}_{D_{j}}^{-1/2} = \mathbf{F}_{D_{j}}\mathbf{\Lambda}_{D_{j}}\bar{\mathbf{F}}_{D_{j}}^{H}$.

To solve the non-convex problem in the form of a difference of concave functions as in \eqref{Diff_Con}, we use the minorization maximization (MM) algorithm
\cite{Yuille03CCCP}\cite{Hunter04MM},
which applies an affine approximation at every iteration step. Using the first-order approximation of a concave function $u(y) \approx u(x) + \nabla u(x)(y-x)$, $h(\mathbf{Q}_{DU})$ at the $n$-th iteration is approximated as
\begin{align}\label{First_Approx}
 h\left(\mathbf{Q}_{DU}\right) \approx h\left(\mathbf{Q}_{DU}^{(n)}\right) + \nabla h\left(\mathbf{Q}_{DU}^{(n)}\right)\left(\mathbf{Q}_{DU} - \mathbf{Q}_{DU}^{(n)}\right).
\end{align} Then, at the $n$-th iteration, the objective function $f\left(\mathbf{Q}_{U},\mathbf{Q}_{DU}\right)$ is lower bounded as
\begin{align}\label{Minorization}
    \tilde{f}^{(n)}\left(\mathbf{Q}_{U},\mathbf{Q}_{DU}\right) &= g\left(\mathbf{Q}_{U},\mathbf{Q}_{DU}\right) - h\left(\mathbf{Q}_{DU}^{(n)}\right) - \nabla h\left(\mathbf{Q}_{DU}^{(n)}\right)\left(\mathbf{Q}_{DU}-\mathbf{Q}_{DU}^{(n)}\right) \nonumber \\
    &=  g\left(\mathbf{Q}_{U},\mathbf{Q}_{DU}\right)  - h^{(n)}\left(\mathbf{Q}_{DU}\right)\nonumber \\
    & \leq f\left(\mathbf{Q}_{U},\mathbf{Q}_{DU}\right)
\end{align} where, from the fact that $\nabla_{\mathbf{x}} \log|\mathbf{I} + \mathbf{AXA}^{H}| = \mathbf{A}^{H}(\mathbf{I} + \mathbf{AXA}^{H})^{-1}\mathbf{A}$,
\begin{align}\label{Approx}
    h^{(n)}\left(\mathbf{Q}_{DU}\right) =& h\left(\mathbf{Q}_{DU}^{(n)}\right) + \nabla h\left(\mathbf{Q}_{DU}^{(n)}\right)\left(\mathbf{Q}_{DU}-\mathbf{Q}_{DU}^{(n)}\right) \nonumber \\
    =& \log \left| \mathbf{I} + \sum_{j=1}^{K_{D}}\mathbf{\bar{G}}_{D_{j}}\mathbf{Q}_{DU_{j}}^{(n)}\mathbf{\bar{G}}_{D_{j}}^{H} \right|  \nonumber \\
   \quad \quad &+ \sum_{j=1}^{K_{D}}\mathrm{Tr}\left( \mathbf{\bar{G}}_{D_{j}}^{H} \left( \mathbf{I} +
        \sum_{k=1}^{K_{D}}\mathbf{\bar{G}}_{D_{k}}\mathbf{Q}_{DU_{k}}^{(n)}\mathbf{\bar{G}}_{D_{k}}^{H}\right)^{-1}\mathbf{\bar{G}}_{D_{j}}
        \left( \mathbf{Q}_{DU_{j}}-\mathbf{Q}_{DU_{j}}^{(n)} \right) \right).
\end{align} Note that  $\tilde{f}$ coincides with $f$ at $(\mathbf{Q}_{U},\mathbf{Q}_{DU})$,
i.e., $f(\mathbf{Q}_{U},\mathbf{Q}_{DU}) = \tilde{f}(\mathbf{Q}_{U},\mathbf{Q}_{DU})$.

Based on the lower bound of the objective function, problem (\ref{New_Sumrate}) is replaced by
\begin{align}\label{New_Sumrate2}
    \underset{ \{\mathbf{Q}_{U_{i}}\}, \{\mathbf{Q}_{DU_{j}}\} }{\mathrm{max}} \quad & \tilde{f}^{(n)}\left(\mathbf{Q}_{U},\mathbf{Q}_{DU}\right) = g\left(\mathbf{Q}_{U},\mathbf{Q}_{DU}\right) - h^{(n)}\left(\mathbf{Q}_{DU}\right)   \nonumber \\
    \mathrm{subject~to} \quad & \mathrm{Tr} \left(\mathbf{Q}_{U_{i}}\right) \leq P_{U_{i}}, \quad i=1,\ldots,K_{U} \nonumber \\
                                & \sum_{j=1}^{K_{D}}\mathrm{Tr} \left( \mathbf{Q}_{DU_{j}} \right) \leq P_{D}  \nonumber  \\
                            & \mathbf{Q}_{U_{i}} \succeq 0,~\mathbf{Q}_{DU_{j}} \succeq 0, \quad i=1,\ldots,K_{U},~j=1,\ldots,K_{D}.
\end{align}
%If we denote the solution set of \eqref{New_Sumrate2} at the $n$-th iteration as $(\mathbf{Q}_{U}^{(n)},\mathbf{Q}_{DU}^{(n)})$, the value of $f$ monotonically increases at each iteration step as
%\begin{equation}
%    f\left(\mathbf{Q}_{U}^{(n+1)},\mathbf{Q}_{DU}^{(n+1)}\right) = \tilde{f}^{(n+1)}\left(\mathbf{Q}_{U}^{(n+1)},\mathbf{Q}_{DU}^{(n+1)}\right)
%     \geq \tilde{f}^{(n)}\left(\mathbf{Q}_{U}^{(n)},\mathbf{Q}_{DU}^{(n)}\right) = f\left(\mathbf{Q}_{U}^{(n)},\mathbf{Q}_{DU}^{(n)}\right).
%\end{equation}
Since this alternative problem is a concave problem, the solution can be obtained by the cvx program \cite{CVX13} iteratively until  $\tilde{f}$ converges such that
\begin{equation}\label{Minor_Algorithm}
    \tilde{f}^{(n+1)}\left(\mathbf{Q}_{U},\mathbf{Q}_{DU}\right) \in
    \max_{\mathbf{Q}_{U}\, \mathbf{Q}_{DU}}\tilde{f}^{(n)}\left(\mathbf{Q}_{U},\mathbf{Q}_{DU}\right).
\end{equation} The MM algorithm is summarized in Algorithm 1.

\begin{algorithm}[!t]
\caption{MM algorithm}
\label{Alg:DC}
\begin{algorithmic}[1]
\vspace{0.2in}
    \STATE Transform the downlink channel to the dual uplink channel.
    \STATE Initialize $\mathbf{Q}_{U_{i}}^{(0)}$ for $i=1,\ldots,K_{U}$ and $\mathbf{Q}_{DU_{j}}^{(0)}$ for $j=1,\ldots,K_{D}$;
    $\mathrm{Tr} \left( \mathbf{Q}_{U_{i}}^{(0)} \right) \leq P_{U_{i}}$ and $\sum_{j=1}^{K_{D}} \mathrm{Tr} \left( \mathbf{Q}_{DU_{j}}^{(0)} \right) \leq P_{D}$.

    \REPEAT
    \STATE Solve (\ref{New_Sumrate2}) to find solutions $\mathbf{Q}_{U_{i}}^{\star}$ for $i=1,\ldots,K_{U}$,
    and $\mathbf{Q}_{DU_{j}}^{\star}$ for $j=1,\ldots,K_{D}$.

    \STATE $n=n+1$.
    \STATE Update $\mathbf{Q}_{U_{i}}^{(n)} = \mathbf{Q}_{U_{i}}^{\star}$ for $i=1,\ldots,K_{U}$ and $\mathbf{Q}_{DU_{j}}^{(n)} = \mathbf{Q}_{DU_{j}}^{\star}$ for $j=1,\ldots,K_{D}$.

    \UNTIL{the sum rate converges.}
    \STATE Transform $\mathbf{Q}_{DU_{j}}^{(n)}$ to $\mathbf{Q}_{D_{j}}$ for $j=1,\ldots,K_{D}$.
\vspace{0.2in}
\end{algorithmic}
\end{algorithm}
%=========================================================================%

%=========================================================================%
\subsection{Alternating Iterative Water-filling Algorithm}
%=========================================================================
Although the MM algorithm is useful to address the non-convex problem of sum rate maximization, it relies on the affine approximation and thus suffers from a degradation of the achievable sum rate due to the approximation errors. Therefore, we develop an algorithm without any approximation to solve the non-convex problem of sum rate maximization.

% Algorithm 2
In MAC, it is known that the iterative water-filling algorithm \cite{Yu04Iterative, Jindal05Sum} yields the optimal transmit covariance matrices of the uplink users. Because the reformulated problem is constituted by two coupled multiple access channels (i.e., the original uplink channel and the dual uplink channel), the iterative water-filling algorithm can be a fundamental framework to find a solution. Since the two multiple access channels are coupled, the covariance matrices in the two multiple access channels are required to be jointly updated based on the iterative water-filing algorithm. However, contrary to conventional MAC or dual MAC, the problem is not convex.
Moreover,  the water-filling algorithm for the uplink users differs from that for the dual uplink users by the power constraints, i.e., individual power constraints for the uplink users and a sum power constraint for the dual uplink users, but the problem is a mixture of both.  Therefore, we take an approach of alternately solving two sub-problems, the sum rate maximization of the uplink channel and the sum rate maximization of the dual uplink channel, based on the iterative water-filling algorithm.

First, to derive the covariance matrix $\mathbf{Q}_{U_{i}}$ of uplink user $U_{i}$, all the other covariance matrices, i.e., $\{\mathbf{Q}_{U_{k}}: k \neq i \}$ and $\{\mathbf{Q}_{D_{j}}\}$, are considered fixed. Then, $\mathbf{Q}_{U_{i}}^{(n)}$ at the $n$-th iteration of the algorithm can be obtained by solving the following sub-problem:
\begin{align}\label{SubProb1}
     \underset{ \mathbf{Q}_{U_{i}} }{\mathrm{max}} \quad & \log{\left| \mathbf{I}+\tilde{\mathbf{H}}_{U_{i}}^{(n)}\mathbf{Q}_{U_{i}}\left(\tilde{\mathbf{H}}_{U_{i}}^{(n)}\right)^{H}\right|} + C_{U_{i}}^{(n)} \nonumber \\
    \mathrm{subject~to} \quad & \mathrm{Tr} \left(\mathbf{Q}_{U_{i}}\right) \leq P_{U_{i}}, \quad \mathbf{Q}_{U_{i}} \geq 0~\quad i=1,\ldots,K_{U}
\end{align}
where
\begin{align}
\tilde{\mathbf{H}}_{U_{i}}^{(n)} = \left( \mathbf{I}+\sum_{j=1}^{K_{D}}\mathbf{G}_{D_{j}}\mathbf{Q}_{D_{j}}\mathbf{G}_{D_{j}}^{H}
+ \sum_{k=1}^{i-1}\mathbf{H}_{U_{k}}\mathbf{Q}_{U_{k}}^{(n)}\mathbf{H}_{U_{k}}^{H}+\sum_{k=i+1}^{K_{U}}\mathbf{H}_{U_{k}}\mathbf{Q}_{U_{k}}^{(n-1)}\mathbf{H}_{U_{k}}^{H} \right) ^{-\frac{1}{2}} \mathbf{H}_{U_{i}} \quad \mathrm{and} \nonumber
\end{align}
\begin{align}
C_{U_{i}}^{(n)} &= \log \left|\mathbf{I}+\sum_{j=1}^{K_{D}}\mathbf{G}_{D}\mathbf{Q}_{D_{j}}\mathbf{G}_{D}^{H}
+ \sum_{k=1}^{i-1}\mathbf{H}_{U_{k}}\mathbf{Q}_{U_{k}}^{(n)}\mathbf{H}_{U_{k}}^{H}+\sum_{k=i+1}^{K_{U}}\mathbf{H}_{U_{k}}\mathbf{Q}_{U_{k}}^{(n-1)}\mathbf{H}_{U_{k}}^{H}  \right|\nonumber\\
&\quad  + \log \left|\mathbf{I} + \sum_{j=1}^{K_{D}}\mathbf{\bar{H}}_{D_{j}}^{H}\mathbf{Q}_{DU_{j}}^{(n-1)}\mathbf{\bar{H}}_{D_{j}}\right|
- \log\left|\mathbf{I}+\sum_{j=1}^{K_{D}}\mathbf{G}_{D}\mathbf{Q}_{D_{j}}\mathbf{G}_{D}^{H}\right|. \nonumber
\end{align}
Note that at the $n$-th iteration for the uplink channel, the covariance matrix for the downlink channel $\mathbf{Q}_{D_{j}}$ in the SI term can be obtained from $\mathbf{Q}_{DU_{j}}^{(n-1)}$ using MAC-BC dualtiy. Applying the singular value decomposition (SVD) of
$\tilde{\mathbf{H}}_{U_{i}}^{(n)}(\tilde{\mathbf{H}}_{U_{i}}^{(n)})^{H}$, we obtain
a unitary matrix $\mathbf{U}_{U_{i}}$ and a diagonal matrix $\mathbf{D}_{U_{i}}$ such that
$\tilde{\mathbf{H}}_{U_{i}}^{(n)}(\tilde{\mathbf{H}}_{U_{i}}^{(n)})^{H} = \mathbf{U}_{U_{i}}\mathbf{D}_{U_{i}}\mathbf{U}_{U_{i}}^{H}$. In this way, the covariance matrix of uplink user ${U_{i}}$ is determined as
\begin{equation}
    \mathbf{Q}_{U_{i}}^{(n)} = \mathbf{U}_{U_{i}}\mathbf{\Lambda}_{U_{i}}\mathbf{U}_{U_{i}}^{H}.
\end{equation}
The optimal values of $\mathbf{\Lambda}_{U_{i}}$ are determined by water-filling as
$\mathbf{\Lambda}_{U_{i}} = [ \mu\mathbf{I}- (\mathbf{D}_{U_{i}} ) ^{-1} ]^{+}$
where the operation $ [\mathbf{A} ]^{+}$ denotes a component-wise maximum with zero and
the water-filling level $\mu$ is chosen such that
$\mathrm{Tr} (\mathbf{\Lambda}_{U_{i}} ) \leq P_{U_{i}}$. The procedure is repeated until the covariances for all the uplink users $\{i=1,\ldots,K_{U}\}$ are obtained.

Next, we perform the iterative water-filling algorithm for the dual uplink channel. Contrary to the uplink channel, a total sum power constraint is imposed on the dual uplink. Consequently,
the water-filling algorithm is
performed in group instead of individual allocation as in \cite{Jindal05Sum}. Specifically,
to derive the covariance matrix $\mathbf{Q}_{DU_{j}}$ of dual uplink user $DU_{j}$,
$\{\mathbf{Q}_{U_{i}}\}$ and $\{\mathbf{Q}_{D_{k}}: k \neq j \}$
are considered fixed. Then, solving the following sub-problem, we obtain  $\mathbf{Q}_{DU_{j}}^{(n)}$.
\begin{align}\label{SubProb2}
   \underset{ \{\mathbf{Q}_{DU}\} }{\mathrm{max}} \quad & \sum_{j=1}^{K_{D}} \log{\left| \mathbf{I}+ \tilde{\mathbf{H}}_{D_{j}}^{(n)}\mathbf{Q}_{DU_{j}}(\tilde{\mathbf{H}}_{D_{j}}^{(n)})^{H}\right|}
   + C_{DU_{j}}^{(n)} \nonumber \\
    \mathrm{subject~to} \quad & \sum_{j=1}^{K_{D}}\mathrm{Tr} ( \mathbf{Q}_{DU_{j}} ) \leq P_{D}, \quad  \mathbf{Q}_{DU_{j}} \geq 0,~~~j=1,\ldots,K_{D}.
\end{align}
where
\begin{align}
\tilde{\mathbf{H}}_{D_{j}}^{(n)} = \left( \mathbf{I} + \sum_{k=1,k \neq j}^{K_{D}}\mathbf{\bar{H}}_{D_{k}}^{H}\mathbf{Q}_{DU_{k}}^{(n-1)}\mathbf{\bar{H}}_{D_{k}} \right)^{-\frac{1}{2}}\mathbf{\bar{H}}_{D_{j}}^{H} \quad \mathrm{and} \nonumber
\end{align}
\begin{align}
C_{DU_{j}}^{(n)} &= \log \left|\mathbf{I} + \sum_{k=1,k \neq j}^{K_{D}}\mathbf{\bar{H}}_{D_{k}}^{H}\mathbf{Q}_{DU_{k}}^{(n-1)}\mathbf{\bar{H}}_{D_{k}} \right| \nonumber \\
&\quad + \log \left|\mathbf{I}+\sum_{j=1}^{K_{D}}\mathbf{G}_{D}\mathbf{Q}_{D_{j}}\mathbf{G}_{D}^{H} + \sum_{i=1}^{K_{U}}\mathbf{H}_{U_{i}}\mathbf{Q}_{U_{i}}^{(n)}\mathbf{H}_{U_{i}}^{H} \right|
- \log \left|\mathbf{I}+\sum_{j=1}^{K_{D}}\mathbf{G}_{D}\mathbf{Q}_{D_{j}}\mathbf{G}_{D}^{H}\right|. \nonumber
\end{align}
Likewise, the covariance matrix for the downlink channel $\mathbf{Q}_{D_{j}}$ in the SI term is also transformed from $\mathbf{Q}_{DU_{j}}^{(n-1)}$, using the MAC-BC duality.
Using SVD of $\tilde{\mathbf{H}}_{D_{j}}^{(n)}(\tilde{\mathbf{H}}_{D_{j}}^{(n)})^{H}$ such that
$\tilde{\mathbf{H}}_{D_{j}}^{(n)}(\tilde{\mathbf{H}}_{D_{j}}^{(n)})^{H} = \mathbf{U}_{D_{j}}\mathbf{D}_{D_{j}}\mathbf{U}_{D_{j}}^{H}$,
the transmit covariance matrices are obtained as
\begin{equation}
    \bar{\mathbf{Q}}_{DU_{j}}^{(n)} = \mathbf{U}_{D_{j}}\mathbf{\Lambda}_{D_{j}}\mathbf{U}_{D_{j}}^{H}, \quad j=1,\ldots,K_{D}
\end{equation}
where $\mathbf{\Lambda}_{D_{j}} = [ \nu\mathbf{I} - (\mathbf{D}_{D_{j}}) ^{-1} ]^{+}$
and $\nu$ is chosen to satisfy the total power constraint
$\sum_{j=1}^{K_{D}}\mathrm{Tr} ( \mathbf{\Lambda}_{D_{j}} ) \leq P_{D}$. To guarantee convergence of the iterative water-filling algorithm with a total sum power constraint,  as in \cite{Jindal05Sum}, the covariance matrix at the $n$-th iteration is updated as
\begin{equation}\label{eq:cov}
    \mathbf{Q}_{DU_{j}}^{(n)} = \frac{1}{K_{D}}\bar{\mathbf{Q}}_{DU_{j}}^{(n)} + \frac{K_{D}-1}{K_{D}}\mathbf{Q}_{DU_{j}}^{(n-1)}, \quad j=1,\ldots,K_{D}
\end{equation}
which ensures the non-decreasing property. Finally, using the dual uplink-downlink transformation, the covariance matrices for the dual uplink channel is transformed
to the covariance matrices of the original downlink channel.

With the covariance matrices obtained at the $n$-th iteration, subproblems in (\ref{SubProb1}) and (\ref{SubProb2}) are sequentially solved to obtain the  covariance matrices at the $(n+1)$-th iteration. This procedure is repeated until the sum rate objective converges. The proposed alternating algorithm is summarized in Algorithm 2 and its convergence is proved in the following theorem.

%=========================================================================%
\begin{algorithm}[!t]
\caption{Alternating IWF algorithm}
\label{Alg:IWF_FD}
\begin{algorithmic}[1]
\vspace{0.2in}
    \STATE Transform the downlink channel to the dual uplink channel.
    \STATE Initialize $\mathbf{Q}_{U_{i}}^{(0)}$ for $i=1,\ldots,K_{U}$ and $\mathbf{Q}_{DU_{j}}^{(0)}$ for $j=1,\ldots,K_{D}$;
    $\mathrm{Tr} \left( \mathbf{Q}_{U_{i}}^{(0)} \right) \leq P_{U_{i}}$ and $\sum_{j=1}^{K_{D}} \mathrm{Tr} \left( \mathbf{Q}_{DU_{j}}^{(0)} \right) \leq P_{D}$.
    \STATE Transform $\mathbf{Q}_{DU_{j}}^{(0)}$ to $\mathbf{Q}_{D_{j}}$, $\forall j=1,\ldots,K_{D}$.
    \STATE Set $n=1$.

    \REPEAT
    \FOR{$i=1$ to $K_{U}$}
    \STATE Solve (\ref{SubProb1}) using the water-filling algorithm to find $\mathbf{Q}_{U_{i}}^{(n)}$
    while keeping all other variables fixed.
    \ENDFOR

    \FOR{$j=1$ to $K_{D}$}
    \STATE Calculate SVD of $\tilde{\mathbf{H}}_{D_{j}}^{(n)}\left(\tilde{\mathbf{H}}_{D_{j}}^{(n)}\right)^{H}$
    for water-filling while keeping all other variables fixed.
    \ENDFOR

    \FOR{$j=1$ to $K_{D}$}
    \STATE Solve (\ref{SubProb2}) using the water-filling algorithm to find $\mathbf{Q}_{DU_{j}}^{(n)}$.
    \ENDFOR

    \STATE Transform $\mathbf{Q}_{DU_{j}}^{(n)}$ to $\mathbf{Q}_{D_{j}}$, $\forall j=1,\ldots,K_{D}$.
    \STATE $n=n+1$.
    \UNTIL{the sum rate converges.}
\vspace{0.2in}
\end{algorithmic}
\end{algorithm}
%=========================================================================%

% Convergence of algorithm 2
\begin{theorem}
Algorithm 2 converges for any $K_{U}$ and $K_{D}$.
\end{theorem}
\begin{IEEEproof}
Algorithm 2 is based on alternating iterations between two subproblems for the uplink channel and the dual uplink channel. Thus, convergence is proved by showing that each step of iteration is non-decreasing in sequence:
\begin{align}
f \left(\mathbf{Q}_{U}^{(n-1)},\mathbf{Q}_{DU}^{(n-1)} \right)
&\leq f \left(\mathbf{Q}_{U}^{(n)},\mathbf{Q}_{DU}^{(n-1)} \right) \label{Eq:1} \\
&\leq f \left(\mathbf{Q}_{U}^{(n)},\mathbf{\bar{Q}}_{DU}^{(n)} \right) \label{Eq:2} \\
&\leq f\left(\mathbf{Q}_{U}^{(n)}, \mathbf{Q}_{DU}^{(n)}\right) .\label{Eq:3}
\end{align}

First, (\ref{Eq:1}) holds since for  fixed $\mathbf{Q}_{DU}^{(n-1)} $, the update from
$\mathbf{Q}_{U}^{(n-1)}$ to $\mathbf{Q}_{U}^{(n)}$ is made by solving subproblem  (\ref{SubProb1}) with the iterative water-filling algorithm,  which corresponds to conventional MAC and thus the iterative water-filling algorithm ensures a non-decreasing update \cite{Yu04Iterative}.

Second, \eqref{Eq:2} is due to the sum power iterative water-filling algorithm with an additional averaging step, for given $\mathbf{Q}_{U}^{(n)}$. Specifically, define an expanded function \cite{Jindal05Sum} as
\begin{align}
f^{exp}\left( \mathbf{Q}_{U}, \mathbf{S}(1), \ldots,\mathbf{S}(K_{D}) \right)
= R_{U} + \frac{1}{K_{D}}\sum_{k=1}^{K_{D}}\log \left| \mathbf{I} + \sum_{j=1}^{K_{D}} \mathbf{\bar{H}}_{D_{j}}^{H}\mathbf{S}\left([j-k+1]_{K_{D}}\right)_{j}\mathbf{\bar{H}}_{D_{j}} \right|,
\end{align}where  $\mathbf{\bar{Q}}_{DU_j}=\mathbf{S}(l)_{j}$  for any $l \in \{1,\ldots,K_{D}\}$, $\mathbf{\bar{Q}}_{DU_j}=\frac{1}{K_{D}}\sum_{k=1}^{K_{D}}\mathbf{S}([j-k+1]_{K_{D}})_{j}$ and $[x]_{K_D}=\mathrm{mod}(x,K_{D})$.
Note that due to the concavity of $\log|\cdot|$, we always have
\begin{align}
f \left( \mathbf{Q}_{U}, \mathbf{\bar{Q}}_{DU} \right) \geq f^{exp}\left( \mathbf{Q}_{U}, \mathbf{S}(1), \ldots,\mathbf{S}(K_{D}) \right).
\end{align} As a result,  \eqref{Eq:2} is satisfied for the $\mathbf{\bar{Q}}_{DU}^{(n)}$ obtained by
\begin{align}
\mathbf{\bar{Q}}_{DU}^{(n)}
&= \arg \max_{\mathbf{\bar{Q}}_{DU}} f^{exp} \left(\mathbf{Q}_{U}^{(n)}, \mathbf{\bar{Q}}_{DU},\mathbf{Q}_{DU}^{(n-1)}, \ldots, \mathbf{Q}_{DU}^{(n-1)} \right) \nonumber \\
&=\arg \max_{\mathbf{\bar{Q}}_{DU}: \mathbf{\bar{Q}}_{DU_j} \geq 0, \sum_{j=1}^{K_{D}}\mathrm{Tr}(\mathbf{\bar{Q}}_{DU_{j}}) \leq P_{D}}
\sum_{k=1}^{K_{D}} \log{\left| \mathbf{I} + \mathbf{\bar{H}}_{D_{k}}^{H}\mathbf{\bar{Q}}_{DU_{k}}\mathbf{\bar{H}}_{D_{k}}
+ \sum_{j \neq k} \mathbf{\bar{H}}_{D_{j}}^{H}\mathbf{Q}_{DU_{j}}^{(n-1)}\mathbf{\bar{H}}_{D_{j}}\right|}  \nonumber \\
&=\arg \max_{\mathbf{\bar{Q}}_{DU}: \mathbf{\bar{Q}}_{DU_j} \geq 0, \sum_{j=1}^{K_{D}}\mathrm{Tr}(\mathbf{\bar{Q}}_{DU_{j}}) \leq P_{D}}
\sum_{j=1}^{K_{D}} \log{\left| \mathbf{I}+ \tilde{\mathbf{H}}_{D_{j}}^{(n)}\mathbf{\bar{Q}}_{DU_{j}}\left(\tilde{\mathbf{H}}_{D_{j}}^{(n)}\right)^{H}\right|}.
\end{align}

Finally, since the cyclic coordinated algorithm to find the optimal set for the expanded function maximization problem is equivalent to the iterative water-filling algorithm for the dual uplink \cite{Jindal05Sum}, \eqref{Eq:3} follows from
% averaging
\begin{align}\label{Averaging}
&f^{exp}\left(\mathbf{Q}_{U}^{(n)},\mathbf{\bar{Q}}_{DU}^{(n)},\mathbf{Q}_{DU}^{(n-1)},\ldots,\mathbf{Q}_{DU}^{(n-1)} \right)  \nonumber  \\
&= R_{U}^{(n)} + \frac{1}{K_{D}}\sum_{j=1}^{K_{D}} \log \left| \mathbf{I}
 + \mathbf{\bar{H}}_{D_{j}}^{H}\mathbf{\bar{Q}}_{DU_{j}}^{(n)}\mathbf{\bar{H}}_{D_{j}} + \sum_{k=1, k \neq j}^{K_{D}}\mathbf{H}_{D_{k}}^{H}\mathbf{Q}_{DU_{k}}^{(n-1)}\mathbf{H}_{D_{k}}\right| \\
&\leq R_{U}^{(n)} + \log \left| \frac{1}{K_{D}}\sum_{j=1}^{K_{D}}\left(\mathbf{I} + \mathbf{\bar{H}}_{D_{j}}^{H}\mathbf{\bar{Q}}_{DU_{j}}^{(n)}\mathbf{\bar{H}}_{D_{j}}
+ \sum_{k=1,k \neq j}^{K_{D}}\mathbf{\bar{H}}_{D_{k}}^{H}\mathbf{Q}_{DU_{k}}^{(n-1)}\mathbf{\bar{H}}_{D_{k}}\right)\right|  \label{Eq:4} \\
&= R_{U}^{(n)} + \log \left| \mathbf{I} + \sum_{j=1}^{K_{D}}\mathbf{\bar{H}}_{D_{j}}^{H}\left(\frac{1}{K_{D}}\mathbf{\bar{Q}}_{DU_{j}}^{(n)}
+ \frac{K_{D}-1}{K_{D}}\mathbf{Q}_{DU_{j}}^{(n-1)}\right)\mathbf{\bar{H}}_{D_{j}}\right| \\
&= f\left(\mathbf{Q}_{U}^{(n)},\mathbf{Q}_{DU}^{(n)}\right).
\end{align}
where \eqref{Eq:4} is due to  the concavity of the $\log|\cdot|$ function and $\mathbf{Q}_{DU}^{(n)}$ is the set of updated covariance matrices from \eqref{eq:cov}.
\end{IEEEproof}

%

%=========================================================================%
%                                                                     %
%=========================================================================%
\section{Numerical Results}\label{Sec:Results}
%=========================================================================%
In this section, we show the average sum rates achieved by the proposed algorithms and verify their convergence. In addition, we evaluate complexity of the proposed algorithms. For numerical performance evaluations, we consider a single cell environment constituted by a FD BS with $M$ antennas, $K_{U}$ HD uplink users, and  $K_{D}$ HD downlink users. The uplink and downlink users have $N$ antennas each. We assume $K_{U}=K_{D}=4$ and $M=N=4$ unless otherwise stated. The BS and uplink user transmit with power $P_{D}=27~\mathrm{dBm}$ and $P_{U}=20~\mathrm{dBm}$, respectively, when each uplink user has the same transmit power as $P_{U_{i}}=P_{U}$ for $\forall i$. Each element of the uplink and downlink channel matrices is realized as i.i.d complex Gaussian random variables with zero mean and variance $\sigma_D^2$ for downlink and $\sigma_U^2$ for uplink. Also, the elements of the interference channel matrices are i.i.d. complex Gaussian random variables with zero mean and variance of $\sigma_{SI}^{2}$ for the residual SI and $\sigma_{CCI}^{2}$ for the CCI. According to \cite{Bharadia14Full}, SI before cancellation is almost the same as the transmit power at BS but with cancellation  SI can be suppressed up to approximately 110 dB. So we set the SI cancellation capability to be $C_{SI}=110$ dB. Then, given BS transmit power of 27 dBm, the residual SI is assumed to be $\sigma^{2}_{SI}=-83~$dBm. According to the line-of-sight (LOS) path-loss model in \cite{3GPP10TR} given by $L_{LOS}=103.4+24.2\log_{10}d$, the path-loss between the BS and a user is assumed to be 91 dB which corresponds to the distance of about $0.3$ km. Thus, for $P_{D}=27~\mathrm{dBm}$ and $P_{U}=20~\mathrm{dBm}$,  we assume $\sigma_D^2=-64$ dBm and $\sigma_U^2=-71$ dBm. The path-loss from an uplink user to a downlink user follows the non-line-of-sight (NLOS) path-loss model in \cite{3GPP10TR} given by $L_{NLOS}=145.4+37.5\log_{10}d$. For $P_{U}=20~\mathrm{dBm}$ and the distance of $0.05$ km, the CCI channel path-loss is assumed to be 97 dB and so $\sigma_{CCI}^{2} = -77$ dBm. To evaluate the effect of interference, the ratio of the received interference power to the desired signal power is defined as $\rho_{SI}$ and $\rho_{CCI}$ for SI at the BS and for CCI at the downlink users, respectively.

%=========================================================================%
\subsection{Convergence}\label{Subsec:Conv}
%=========================================================================%
%=========================================================================%
%\clearpage
\begin{figure}[!t]
    \begin{center}
\centering
\includegraphics[width=12cm]{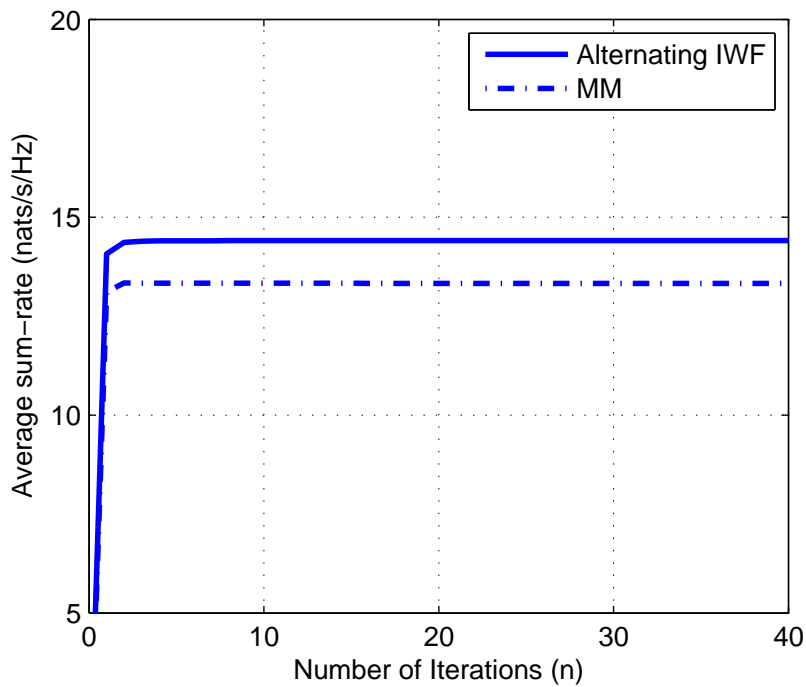}
    \caption{Convergence of the proposed algorithms with $K_{U}=K_{D}=4$ and $M=N=4$}
    \label{Fig:Converge}
    \end{center}
\end{figure}
%\clearpage
%=========================================================================%
For given assumptions, we evaluate convergence rate of the proposed algorithms in Fig. \ref{Fig:Converge}. Both algorithms converge within 3 or 4 iterations, although each algorithm has different computational complexity for each iteration.
%=========================================================================%
\subsection{Sum Rate}
%=========================================================================%

\begin{figure*}[!htp]
\begin{center}
\subfigure[]{
	 \includegraphics[width=11cm]{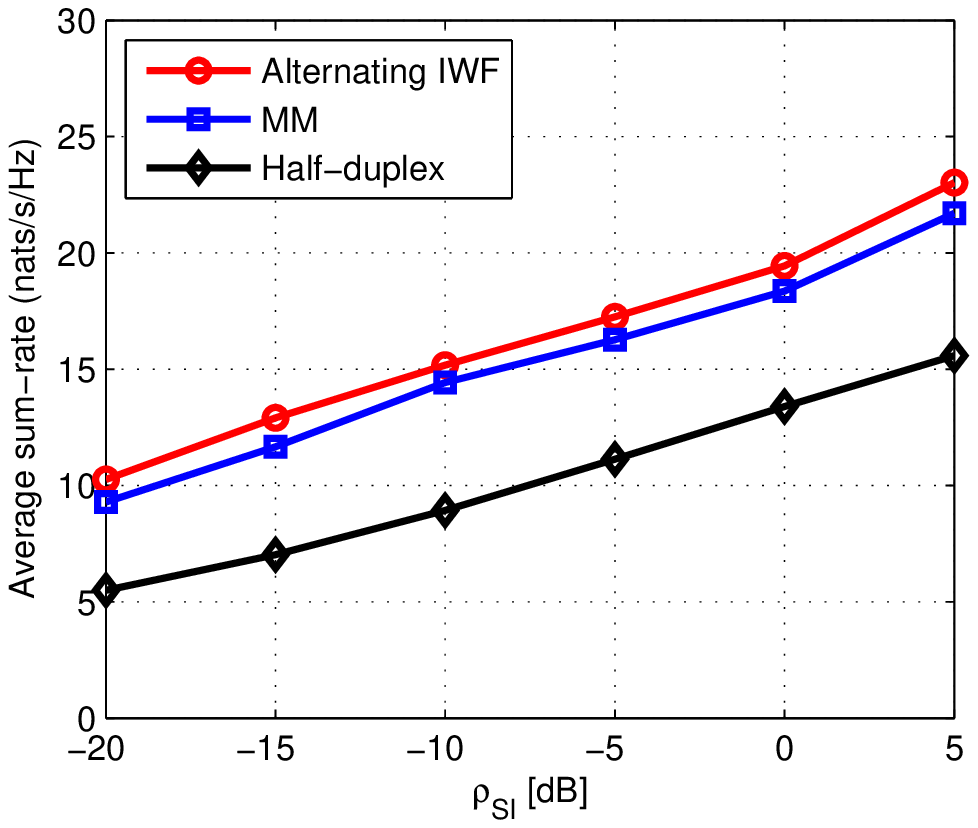}
    	 \label{Fig:Rate_P_D}
}
\subfigure[]{
	\includegraphics[width=11cm]{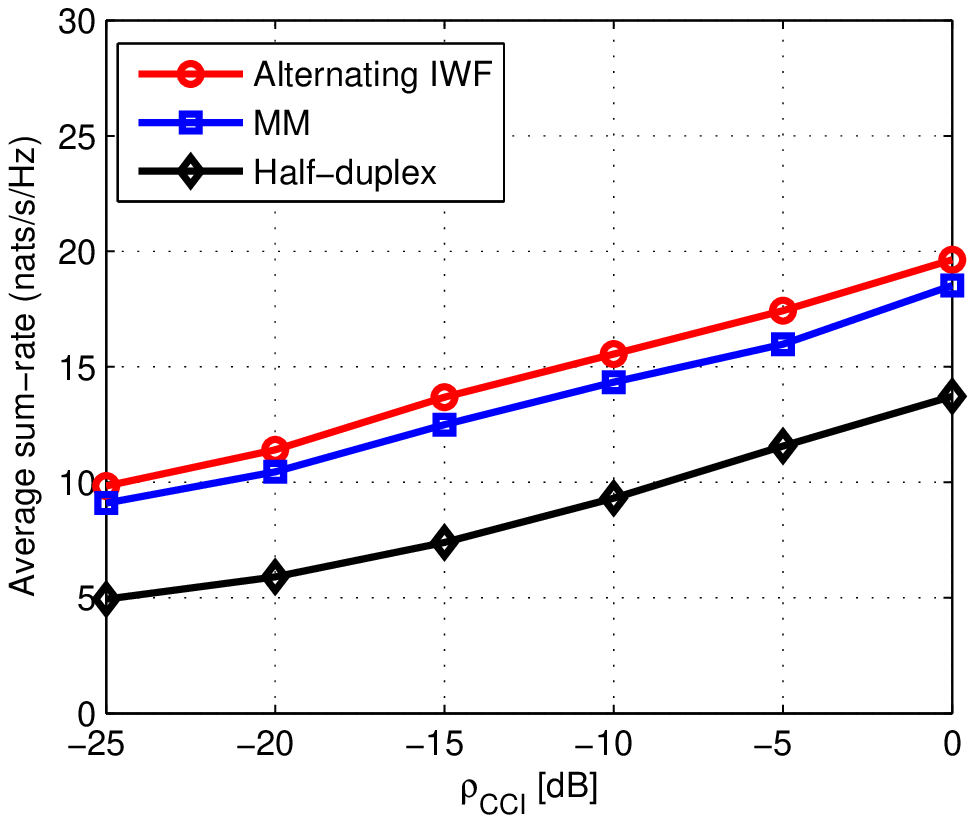}
	\label{Fig:Rate_P_U}
}
\caption{ (a) Average sum rate versus SI with varying transmit power of BS ($P_{D}$)
		(b) Average sum-rate versus CCI with varying transmit power of uplink user ($P_{U}$)}
\label{Rate_Interf1}
\end{center}
\end{figure*}

Fig. \ref{Rate_Interf1} shows the average sum rate versus the ratio of the received interference power to the desired signal power. For fixed SI cancellation capability of $C_{SI}=110~\mathrm{dB}$, residual SI is calculated according to the transmit power of BS, $P_{D}$, when $P_{U}=20$ dBm and $\sigma_{CCI}^{2} = -77$ dBm in Fig. \ref{Fig:Rate_P_D}. Also, when the path-loss between an uplink user and a downlink user is fixed as $L_{CCI}=97~\mathrm{dB}$, CCI varies according to the transmit power of the uplink user when $P_{D}=27~\mathrm{dBm}$ and $\sigma^{2}_{SI}=-83$ dBm in Fig. \ref{Fig:Rate_P_U}. As the transmit power grows, the sum rates of the proposed algorithms increase and the alternating IWF outperforms the MM approach.

\begin{figure*}[!htp]
\begin{center}
\subfigure[]{
	 \includegraphics[width=11cm]{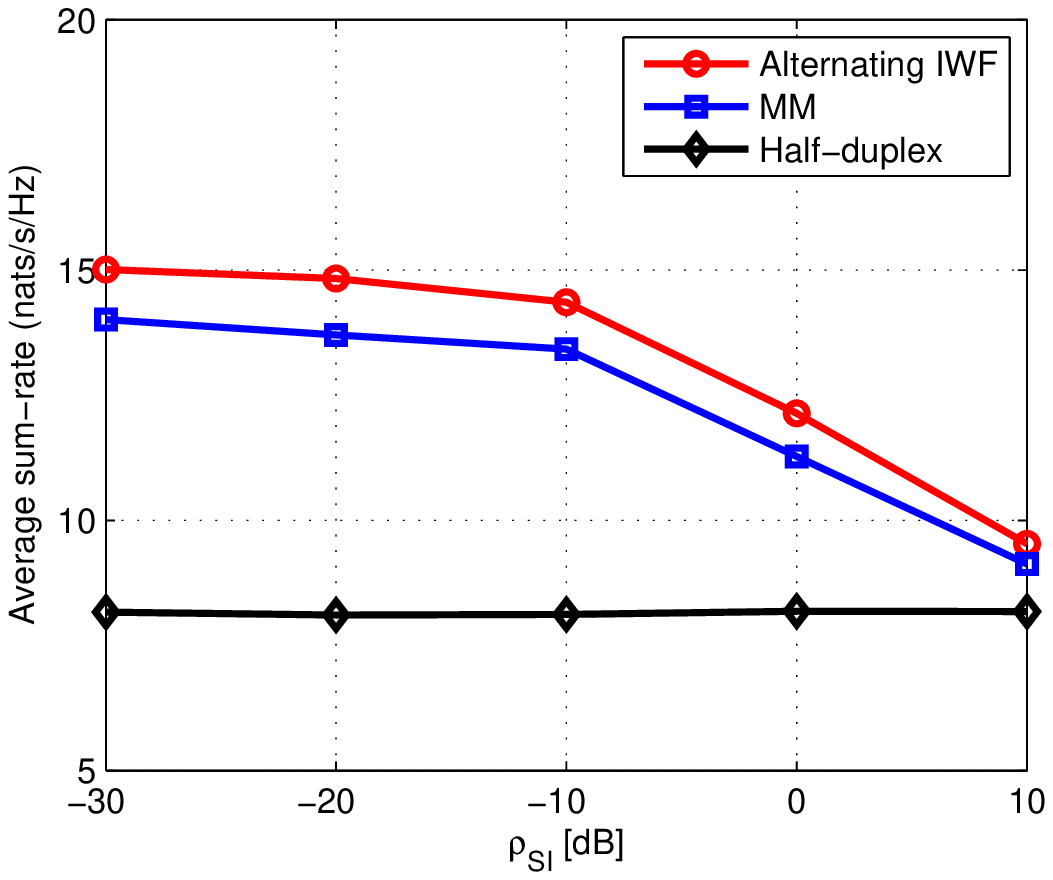}
    	 \label{Fig:Rate_SI}
}
\subfigure[]{
	\includegraphics[width=11cm]{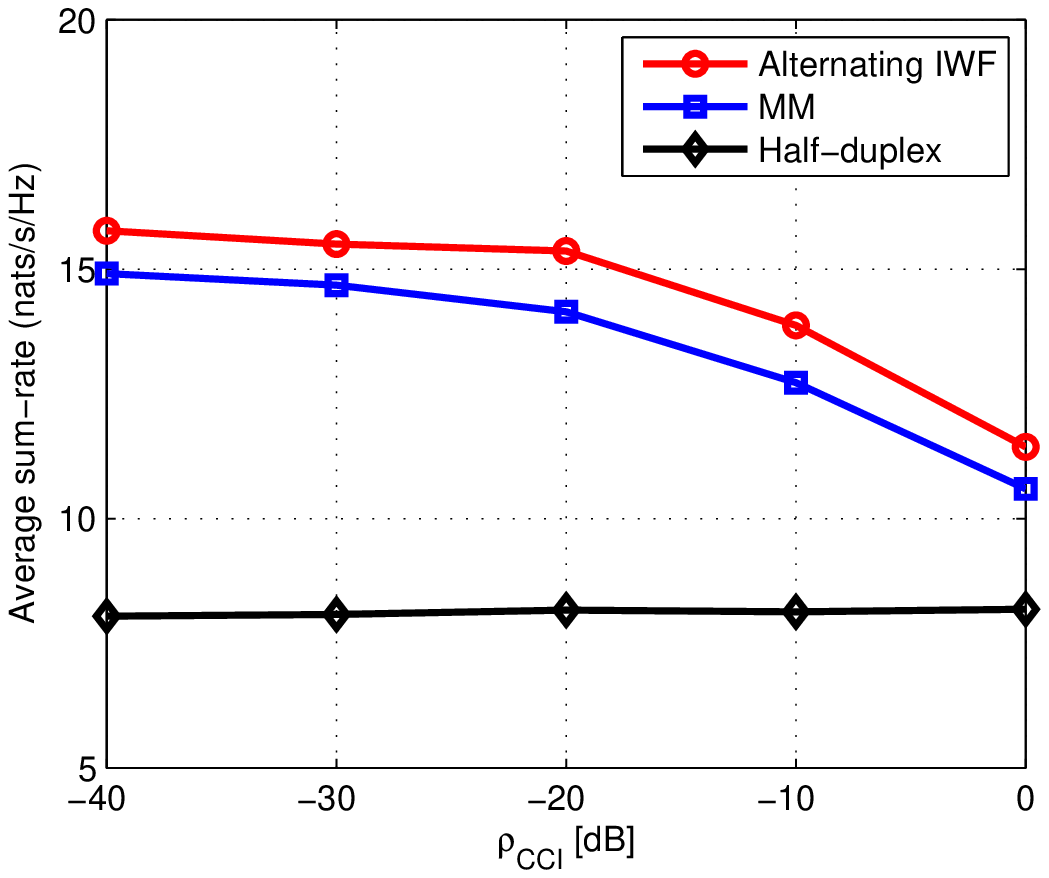}
	\label{Fig:Rate_CCI}
}
\caption{ 	(a) Average sum rate versus SI with varying SI cancellation capability ($C_{SI}$)
		(b) Average sum-rate versus CCI with varying path-loss between uplink and downlink users ($L_{CCI}$) }
\label{Rate_Interf2}
\end{center}
\end{figure*}

To clearly capture the effect of each interference, Fig. \ref{Fig:Rate_SI} exhibits the sum rate versus $\rho_{SI}$ according to SI cancellation capability ($C_{SI}$) when CCI is fixed to be $\sigma_{CCI}^{2} = -77$ dBm, $P_{D}=27$ dBm, and $P_{U}=20$ dBm.  Fig. \ref{Fig:Rate_CCI} shows the sum rate versus $\rho_{CCI}$ defined by path-loss between the uplink and downlink users ($L_{CCI}$) when SI is fixed to be $\sigma_{SI}^{2} = -83$ dBm, $P_{D}=27$ dBm, and $P_{U}=20$ dBm."
For a reference, we also plot the achievable total sum rate of BC and MAC in a HD system with the same numbers of antennas and users as the FD system. In the HD system, the iterative water-filling based algorithm is used for determining the optimal transmit covariance matrices, under individual power constraints \cite{Yu04Iterative} for MAC and a sum power constraint \cite{Jindal05Sum} for BC. The transmit times for BC and MAC are assumed to be the same and the total sum rate of the MAC and BC accounts for the rate loss due to the duplex duty cycle.
In Fig. \ref{Fig:Rate_SI}, as SI grows, the sum rates of both the algorithms decrease since the increased SI degrades the uplink sum rate. Both the algorithms outperform the HD system, owing to the well balanced beamforming, in the presence of CCI (i.e., $\sigma_{CCI}^{2}=-77~\mathrm{dBm}$). In Fig. \ref{Fig:Rate_CCI}, as CCI increases when $\sigma_{SI}^{2}=-83~\mathrm{dBm}$, the sum rate reduces since CCI directly decreases the downlink sum rate. Figs. \ref{Fig:Rate_SI} and \ref{Fig:Rate_CCI} show that the alternating IWF algorithm outperforms the MM algorithm which has a loss from the affine approximation.

%=========================================================================%
\subsection{Complexity}
%=========================================================================%
To analyze computational complexity, we first evaluate the number of floating point operations (FLOP count) required per iteration where either a complex multiplication or a complex addition is counted as one FLOP \cite{Hunger05Floating}.
% MM algorithm
In the MM algorithm, the convex optimization problem in (19) is solved by using cvx solver based on semi-definite programming (SDP). The computational complexity of SDP is obtained by counting the operations of an interior-point method \cite{Luo10Semidefinite}. Thus, computational complexity of solving the problem in (19) is $\mathcal{O}((NK_{U})^{4.5}\log(1/\epsilon))$ for uplink and $\mathcal{O}((NK_{D})^{4.5}\log(1/\epsilon))$ for downlink. As a result, the overall computational complexity of the MM algorithm is $\mathcal{O}(((NK_{D})^{4.5}+(NK_{U})^{4.5})\log(1/\epsilon))$ per iteration where $\epsilon$ is the accuracy target.

% Alternating IWF algorithm
On the other hand, in the alternating IWF algorithm, we can count the exact number of FLOPs at each step. In the uplink, computational complexity is dominated by calculation of an effective channel matrix $\mathbf{\tilde{H}}_{U_{i}}$, eigenvalue decomposition of $\mathbf{\tilde{H}}_{U_{i}}\mathbf{\tilde{H}}_{U_{i}}^{H}$, and  calculation of the covariance matrix $\mathbf{Q}_{U_{i}}$ for $\forall i$. In the downlink,  besides the dominant operations in the uplink, the operation to update the covariance matrix from $\mathbf{\bar{Q}}_{D_{j}}$ to $\mathbf{Q}_{D_{j}}$ has to be additionally taken into account. Consequently, computational complexity for solving the problem in (21) and (23) is $\frac{7}{3}K_{D}M^{3} + \frac{29}{3}K_{U}M^{3} + \frac{4}{3}K_{U}N^{3} + 2K_{U}M^{2}N + 2K_{U}MN^{2}$ and $\frac{29}{3}K_{D}M^{3} + \frac{4}{3}K_{D}N^{3} + 2K_{D}M^{2}N + 2K_{D}MN^{2}$, respectively. Therefore, the total computational complexity of the alternating IWF algorithm is approximately $\mathcal{O}(K_{U}M^{3} + K_{U}N^{3} + K_{D}M^{3} + K_{D}N^{3})$ at each iteration. 

%=========================================================================%
%                                                                     %
%=========================================================================%
\section{Conclusion}\label{Sec:Conclusion}
%=========================================================================%
We developed two iterative algorithms to solve the non-convex problem of sum rate maximization in full-duplex multiuser MIMO systems. Using the MAC-BC duality, we first reformulated the sum rate maximization problem into an equivalent sum rate maximization of MAC to properly address the couple effects of MAC and BC due to self-channel interference and co-channel interference. Although the equivalent problem was still non-convex,  the transformed objective function allowed us to apply the MM algorithm which makes an affine approximation to the difference of concave functions. To avoid performance degradation resulting from the affine approximation, we also devised an alternating algorithm based on iterative water-filling without any approximation to the objective function. The proposed two algorithms were shown to address the coupled design issue well and properly balance between the individual sum rates of MAC and BC to maximize the total sum rate. It was also proved that the proposed two algorithms ensured fast convergence and low computational complexity.

\end{document}